\documentclass[preprint,amsmath,amssymb,nofootinbib]{revtex4}

\usepackage{dcolumn}
\usepackage{bm}
\usepackage{slashed}
\usepackage{kevin}
\usepackage{graphicx}
\usepackage{epstopdf}

\begin{document}
\title{Photons as a viscometer of heavy ion collisions}
\author{Kevin Dusling}
\email{kdusling@quark.phy.bnl.gov}
\affiliation{Physics Department, Building 510A\\Brookhaven National Laboratory\\Upton, NY-11973, USA}

\date{\today}

\begin{abstract}
The viscous correction to thermal photon production at leading log order is calculated and integrated over the space-time evolution of a hydrodynamic simulation of heavy-ion collisions.  The resulting transverse momentum spectra and elliptic flow can be reliably calculated within a hydrodynamic framework up to transverse momenta of $q_\perp\approx 2.5$ GeV and $q_\perp\approx 1.5$ GeV respectively.  A non-vanishing viscosity leads to a larger thermalization time when extracted from the experimentally measured inverse slope ($T_{\mbox{eff}}$) of photon $q_\perp$ spectra.  A precise, $\mathcal{O}($20 MeV$)$, measurement of photon $T_{\mbox{eff}}$ can place stringent bounds on $\tau_0$ and $\eta/s$.

\end{abstract}

\maketitle

\section{Introduction}

There is a general consensus that the matter produced at RHIC behaves as a near perfect fluid \cite{perfectfluid}.  One of the key findings that led to this interpretation is the strong anisotropic flow of produced hadrons and its description by ideal hydrodynamic simulations \cite{idealhydro}.  Although it is too early to draw definitive conclusions it appears that the deviations from ideal hydrodynamic behavior may be ascribed to dissipative effects.  This has been suggested in many of the recent works on dissipative hydrodynamics \cite{vishydro}. 

Care must be taken when drawing conclusions about the viscosity of the early matter at RHIC based on hadronic observables alone.  As hadrons interact strongly, they are sensitive to the later stages of the evolution leaving an ambiguity about whether the viscous effects seen in spectra stem from the hadronic or QGP phase.  On the other hand, electromagnetic probes are emitted throughout the entire space-time evolution reaching the detector without any final state interactions and are not sensitive to the dynamics of freeze-out.

As the next generation of experiments shift towards a precision study of the matter produced at RHIC it is imperative to have multiple observables constraining the medium properties inferred from the data.  This work demonstrates that direct photons can be used to constrain the shear viscosity.  It is also shown, that by neglecting the presence of viscosity, incorrect thermalization times will be extracted from experiments.  

There is a long history of photon calculations that we can't summarize here.  Many of the works relevant to experiment \cite{photonphem} have relied on kinetic equilibrium and others \cite{anisdilep} have studied the effects of early momentum-space anisotropies.  Only recently, however, has there been a measurement \cite{:2008fqa} that is precise enough to suggest the presence of an early hot stage of matter. 

\section{Photon Rates with Viscosity}

\begin{figure}[hbtp]
  \vspace{9pt}
  \centerline{\hbox{ \hspace{0.0in}
\includegraphics[scale=1.2]{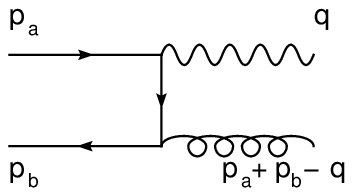}
    \hspace{0.13in}
\includegraphics[scale=1.2]{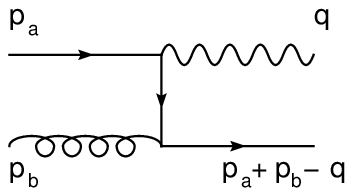}
    }
  }
   \vspace{9pt}
\caption{Feynman diagrams for the annihilation process (left) and the Compton process (right).}
\label{fig:dia}
\end{figure}

In this section we show how the presence of viscosity modifies the photon spectra.  In order to demonstrate the effect we look at the $2\to 2$ processes in fig.~\ref{fig:dia}.  There are additional diagrams that contribute to the thermal photon rate at leading order \cite{Arnold:2002ja}, which will not be examined in this leading log analysis.  The emission rate of photons having momentum $\vec{q}$ and energy $E_q=\vec{q}$ is
\beqa
E_q\frac{dN}{d{\vec q}}&=&4 \sum_f \int\frac{d^3 p_a d^3 p_b}{(2\pi)^6}\nonumber\\
&\times& f_a(p_a) f_b(p_b) \left[1\pm f_2(p_a+p_b-q)\right] E_q\frac{d\sigma}{d{\vec q}} v_{ab}\,,\spc\spc\spc\spc\spc\spc\spc
\eeqa
where the sum is over quark flavors and $f_a(E_a,{\vec p}_a)$ is particle specie $a$'s distribution function, which is not necessarily in equilibrium.  We have used the upper (lower) sign for a final state boson (fermion).

For both the Compton and annihilation process the leading log behavior comes from when the exchanged momentum is soft ({\em i.e.} $ \sim gT$).  In this case the amplitude will be dominated by forward scattering.  Following \cite{WongBk} we can approximate the differential cross section as
\beqa
E_q\frac{d\sigma}{d{\vec q}}\approx E_q \sigma_{tot}(s) \delta^3(\vec{q}-\vec{p}_a)\,,
\eeqa
where $\sigma_{tot}$ is the total cross section of either process.  Performing the integration over ${\vec p}_a$ we are left with
\beqa
E_q\frac{dN}{d{\vec q}}=\frac{2}{(2\pi)^6} f_a(q) \sum_f \int d^3 p_b f_b(p_b) \left[1\pm f_{2}(p_b)\right] \frac{s \sigma(s)}{E_b}\,.
\label{eq:rate1}
\eeqa
By examining the above equation one can already see where the viscous correction will come into play.  The expression for the photon rates is proportional to the distribution function of particle $a$ (the quark).  Therefore, if viscosity modifies the quark's distribution function it will modify the photon emission rates accordingly.  Following \cite{Bellac} we take a one parameter {\em ansatz} for the viscous correction to the distribution function 
\beqa
f^a(p_a) = f_0^a(p_a)+\frac{C_a}{2T^3} f_0^a\left[ 1\pm f_0^a\right] p^\mu_a p^\nu_a \partial_{\langle \mu} u_{\nu \rangle} \,,
\eeqa
where $f_0^a$ is the particle's equilibrium distribution function and $C_a$ is a constant determined in Appendix A and the notation $\langle \cdots \rangle$ designates that the quantity in brackets should be symmetrized and made traceless. 

In general each distribution function in the above rate equation~(\ref{eq:rate1}) must be replaced by its viscous counterpart, $f_0+\delta f$, and then one must drop terms of $\mathcal{O}(\delta f^2)$.  A full analysis, which will be presented elsewhere, has found that the viscous corrections to the distribution functions occurring under the phase space integral lead to corrections to the coefficient under the log.  Therefore, to leading log order, we neglect the viscous correction to $f_b$ and perform the above phase space integrals in the same manner as done in \cite{WongBk, Kapusta:1991qp} where the diverging differential cross section is regulated using the re-summation technique\footnote{In principal one must also include the viscous correction when computing the resummed propagator.  It turns out that these corrections can be taken into account by introducing a generalized momentum dependent thermal mass.  This correction to the rates will not be enhanced by the logarithm and can therefore be neglected in this leading log approach.} of Braaten and Pisarski \cite{Braaten:1989mz}.  

The final result of both the Compton and $\overline{q}q$ annihilation processes is
\beqa
E_q\frac{dN}{d{\vec q}}= \frac{5}{9} \frac{\alpha_e \alpha_s}{2\pi^2} f_a(q) T^2\ln\left[\frac{3.7388 E_q}{g^2 T}\right]\,,
\label{eq:log}
\eeqa
where $f_a$ is the quark's off-equilibrium distribution function
\beqa
f^a(q) = f_0^a(q)+1.3\frac{\eta/s}{2T^3} f_0^a\left[ 1- f_0^a\right] q^\mu q^\nu \partial_{\langle \mu} u_{\nu \rangle} \,.
\eeqa

\section{Spectra in Ultra-relativistic Heavy-Ion Collisions}

In this section the above photon rates are integrated over the space-time evolution of the collision region determined by a 2+1 dimensional boost invariant viscous hydrodynamic model \cite{Dusling:2007gi}.  The same bag model equation of state and Glauber model initial conditions of \cite{Teaney:2001av} are used since it was able to predict many hadronic observables.  We have considered photon production from the QGP phase only.  An impact parameter of $b=6.5$ fm and $\eta/s=1/4\pi$ is used throughout.  The one parameter in the rate equations, $\alpha_s$, is evaluated at the scale $\mu=\pi T$ from the two loop $\beta$ function.  Because we are using the leading-log rates the results below $q_\perp\sim 1$ GeV are speculative.  Above 1 GeV the expression under the log in eq.~\ref{eq:log} remains larger than one.

There is one technicality regarding the viscous evolution model which must be discussed.  We have chosen to use the ideal results for the evolution model for the viscous case as well.  This amounts to neglecting the viscous corrections to the flow and temperature profiles which tend to be small, especially in the early stages of the evolution.  Even though this approach is fundamentally inconsistent (energy-momentum is violated when converting from hydro to particles) the corrections are small.  This procedure is convenient for two reasons. First, arbitrarily large gradients can be treated.  And second, the final multiplicities will remain unchanged so we will not have to worry about modifying the initial conditions in order to re-fit hadronic observables at finite viscosity.  The entire viscous effect, in this work, comes from the modification to the rates as outlined in the previous section.

A previous work \protect\cite{Dusling:2008xj} studied the emission of dileptons from a viscous medium taking into account the viscous correction to the underlying flow and temperature profiles.  While the viscous correction to the hydrodynamic variables modified the yields the shape of the spectrum (as seen through $T_{\mbox{eff}}$) was largely undistorted.  This is further motivation for neglecting the viscous corrections to flow. 

\begin{figure}[hbtp]
  \vspace{2pt}
  \centerline{\hbox{ \hspace{0.0in}
\includegraphics[scale=.35]{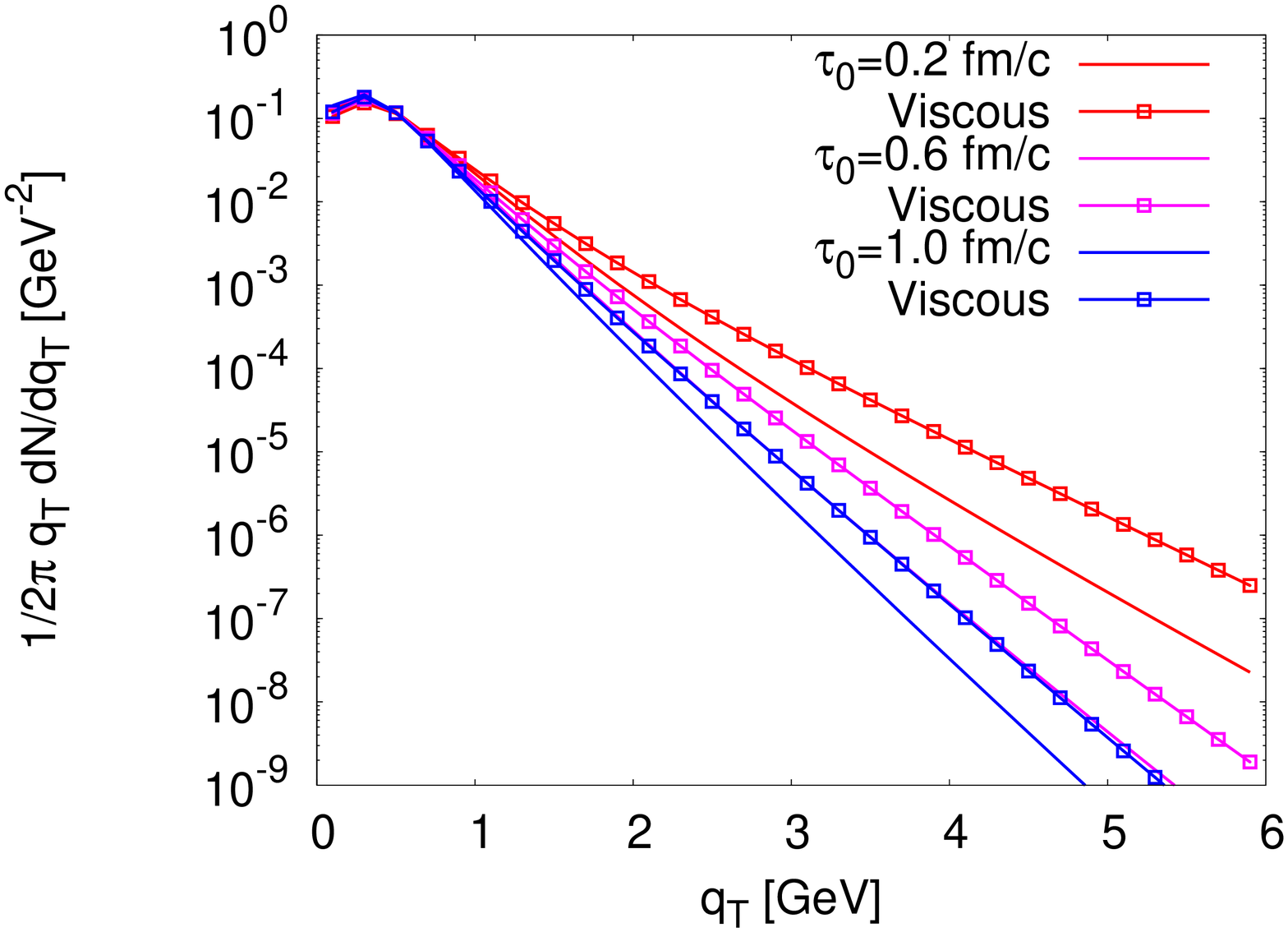}
    }
  }
  \vspace{2pt}
  \centerline{\hbox{ \hspace{0.4in}
\includegraphics[scale=.32]{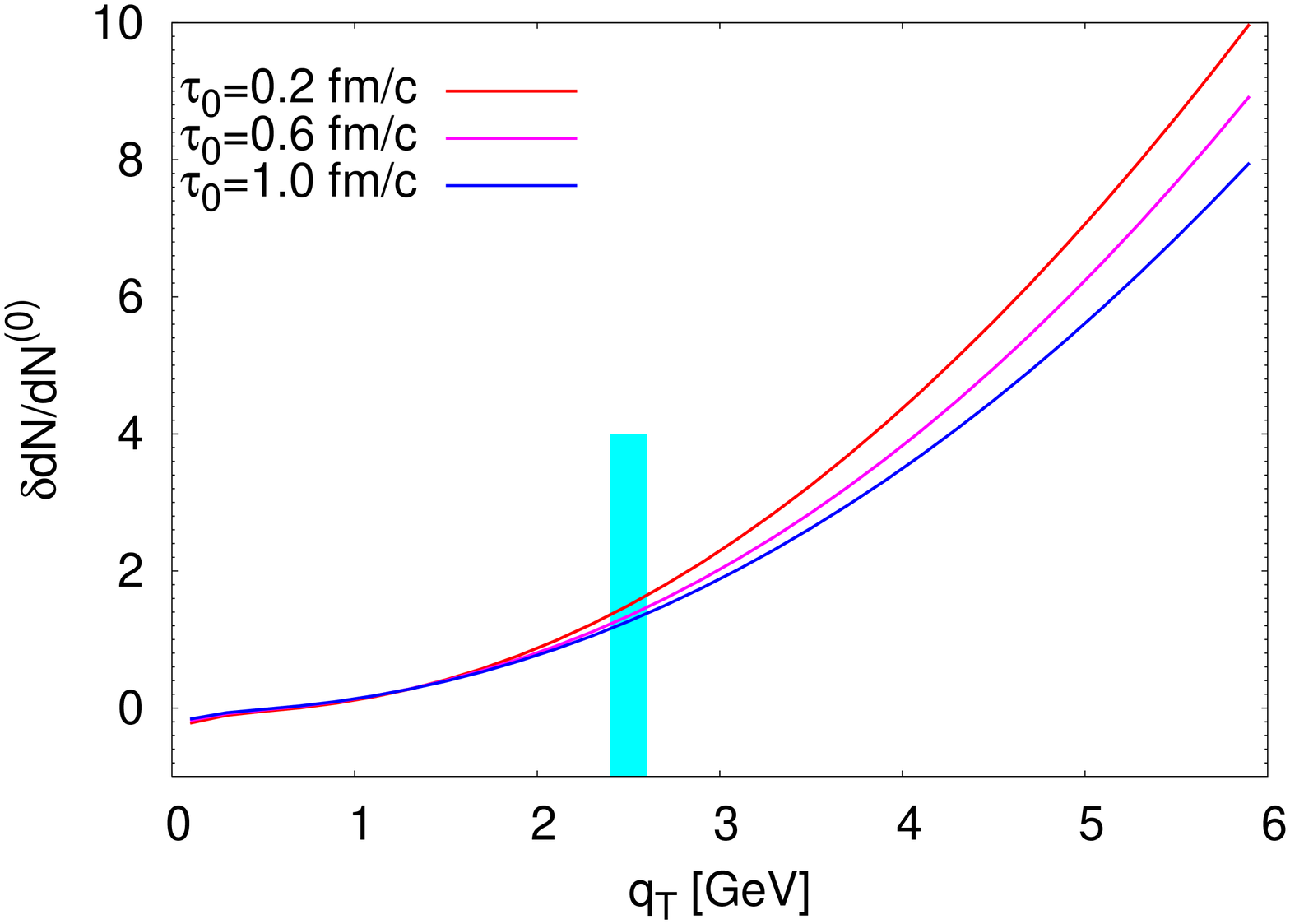}
    }
  }
  \vspace{2pt}
\caption{$q_\perp$ spectra of thermal photons from the QGP (top) and the ratio of the viscous correction over the ideal result (bottom).  The band in the lower figure indicates where the $q_\perp$ spectra can no longer be reliably calculated.}
\label{fig:qt}
\end{figure}

Fig.~\ref{fig:qt} shows the thermal photon transverse momentum spectra for the hydrodynamic model having starting times of $\tau_0=0.2, 0.6, 1.0$ fm/c.  The ideal results are shown as solid lines.  The dominant contribution at higher momentum comes from the early stages of the evolution where the temperature is highest.  This leads to the expected increase in yields at higher momentum for earlier starting times.  Shown as lines with symbols are the corresponding viscous results. A hardening of the transverse momentum spectra is seen, reminiscent of the known effect of viscosity on single particle spectra \cite{Teaney:2003kp}.  These results should not come as a surprise.  At earlier times the collision geometry has larger gradients and the underlying quasi-particles are furthest from local thermal equilibrium.  Viscosity introduces a power law correction to the spectra (reminiscent of particle production in  perturbative QCD).  As the system evolves, thermalization occurs by transferring momentum by \brem or collisions to softer modes until the spectra eventually become thermal.  Since the relaxation time grows with energy incomplete thermalization enhances the quark distribution at high $q_T$.  The harder distribution of quarks leads to a harder spectrum of photons. 
  
The dominate contribution to the transverse momentum spectra at high $q_\perp$ is from the earliest times when the medium is hottest and the gradients are largest\footnote{The viscous correction is proportional to $\partial_\mu u^\mu\sim 1/\tau$ at early times}.  This leads to a larger viscous correction, and the subsequent breakdown of the hydrodynamic description, at large $q_\perp$.  The lower plot of fig.~\ref{fig:qt} shows the ratio of the viscous correction to the ideal result.  The correction becomes of order one at $q_\perp\approx 2.5$ GeV.

\begin{figure}
\includegraphics[scale=.35]{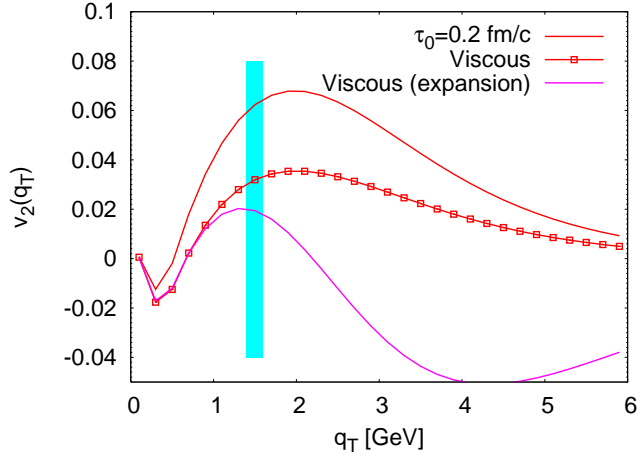}
\caption{Elliptic flow of thermal photons from the QGP.}
\label{fig:v2}
\end{figure}

Fig.~\ref{fig:v2} shows the photon elliptic flow defined as
\beqa
v_2 \equiv \frac{\int d\phi \cos(2\phi)\spc dN +\delta dN}{\int d\phi\spc dN +\delta dN}\approx \frac{\int d\phi \cos(2\phi)\spc dN +\delta dN}{\int d\phi\spc dN} -\frac{\int d\phi \spc \delta dN \int d\phi \cos(2\phi)\spc dN}{(\int d\phi\spc dN)^2}
\label{eq:v2def}
\eeqa
where $dN=dN/d^3q$ is the space-time integrated ideal photon spectra and $\delta dN=\delta dN/d^3q$ is the viscous correction.  In the rightmost expression the denominator has been expanded in order to keep terms up to $\mathcal{O}(\delta f)$ only.  The solid red line shows the ideal result.  As explained in \cite{Chatterjee:2005de}, the total photon $v_2$ follows from a weighted average of the flow over the proper time of the evolution.  A linear rise in $v_2$, as expected from hydro, is observed.  The suppression at large $q_\perp$ is due to the early non-flowing phase, which dominates the yields at large momentum.  Including viscosity, shown as the line with symbols, has a large effect on the $v_2$.  The ideal result is suppressed by a factor as large as $\sim 2$.  In contrast to the hadronic $v_2$, where the largest effect of viscosity is at high $p_\perp$, we find large corrections at {\em all} $q_\perp$ in the case of photons.  

The solid magenta line in Fig.~\ref{fig:v2} shows the viscous result using the expansion in the rightmost expression of eq.~\ref{eq:v2def}.  When the two results disagree this observable can no longer be reliably calculated.  This happens when $q_\perp\approx 1.5$ GeV as shown by the solid band.

\section{Discussion}

\begin{figure}
\includegraphics[scale=.35]{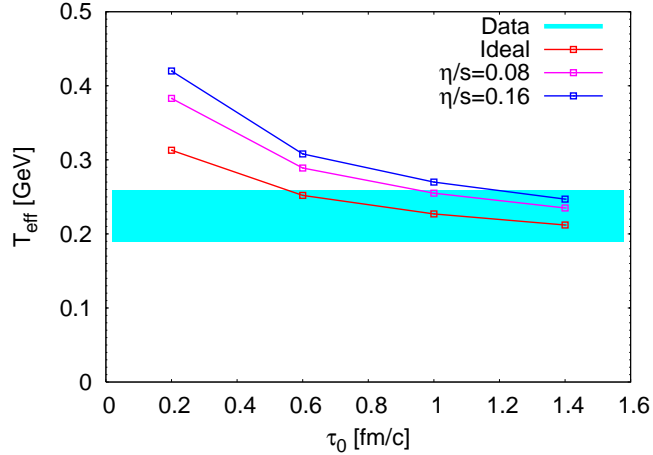}
\caption{Effective temperature of final photon spectra for starting times of $\tau_0=0.2,0.6,1.0$ and 1.4.  Lines are shown to guide the eye.  The solid band is the measured slope from minimum bias data \protect\cite{:2008fqa}.}
\label{fig:phteff}
\end{figure}

The sensitivity of the spectra to the shear viscosity makes the extraction of the thermalization time more subtle.  It also makes it possible to extract $\eta/s$ from the data.  In order to demonstrate the process, the inverse slope of the photon spectra is computed by a fit to $1/q_\perp dN/dq_\perp\propto \exp(-q_\perp/T_{\mbox{eff}})$ in the momentum region\footnote{Realistic model calculations \protect\cite{photonphem} find that the QGP contribution dominates the yields in this kinematic region.} $1.5 \leq q_\perp$ (GeV) $\leq 2.5$.  In fig.~\ref{fig:phteff} the effective temperature $T_{\mbox{eff}}$ is plotted versus the thermalization time $\tau_0$ for $\eta/s=0, 1/4\pi$ and $2/4\pi$.  Both viscosity and earlier thermalization cause the effective temperature to increase in a non-trivial way.  

We should clarify to the reader that what we call {\em thermalization time} is really the hydrodynamic starting time.  In this work any production before $\tau_0$ has been neglected.  If one included non-equilibrium production from the early evolution prior to $\tau_0$ we would expect to see the strong dependence on $\tau_0$ to be reduced. 
 
From fig.~\ref{fig:phteff}, it is clear that fitting the data with ideal hydrodynamic simulations will result in the extraction of earlier thermalization times.  For example, the same $T_{\mbox{eff}}$ is observed for an ideal (viscous) evolution starting at $\tau_0=$0.6 (1.0) fm/c. 

A precise measurement of the inverse slope could constrain a combination of $\tau_0$ and $\eta/s$.  In order to discern between the ideal and viscous results shown here a measurement must pin down $T_{\mbox{eff}}$ to within 20 MeV.   The band in fig.~\ref{fig:phteff} shows the experimentally measured slope \cite{:2008fqa}, including both systematic and statistical errors, in order to demonstrate the current quality of the data. 

There are a number of caveats which must be discussed before a fair comparison can be made with data.  The largest uncertainty comes from using the leading log results.  Examining the full leading order results from \cite{Arnold:2002ja} we estimate that the leading log contribution used here comprises about fifty percent of the thermal QGP photon yields in the kinematic regions of interest.  Therefore, even if the viscous correction to the remaining leading order (non leading log) results turned out to be negligible, one would still expect the above conclusions to hold at a qualitative level.  

\section{Conclusions}
The viscous correction to thermal photons at leading log order is computed.  The rates are then integrated over the space-time history of a hydrodynamic simulation.  The viscosity sets a bound on where a hydrodynamic description is reliable ($q_\perp\sim 2.5$ GeV for spectra and $q_\perp\sim 1.5$ GeV for elliptic flow).  We have shown that viscosity increases $T_{\mbox{eff}}$ and by neglecting its presence wrong conclusions about the thermalization time will be reached.  Our model calculation has shown that the photon spectra can place stringent constraints on both $\tau_0$ and $\eta/s$.

\begin{acknowledgments}
I am grateful to Raju Venugopalan for a careful reading of this manuscript and making many valuable suggestions.  I would also like to thank Ulrich Heinz, Shu Lin, Rob Pisarski, Derek Teaney and Werner Vogelsang for useful discussions.  This work was supported by the US-DOE grant DE-AC02-98CH10886.
\end{acknowledgments}

\appendix
\section{Strength of the off-equilibrium quark distribution}

In this appendix we make an estimate for $C_q$.  From the definition of the viscous part of the stress-energy tensor
\beqa
\delta T^{\mu\nu}\equiv \eta \partial^{\langle \mu} u^{\nu \rangle}=\sum_{a}\nu_a\int \frac{d^3p}{(2\pi)^3}\frac{p^\mu p^\nu}{E_p}p^\alpha p^\beta \partial_{\langle \alpha} u_{\beta \rangle}\delta f^a(p)\,,\nonumber\\
\eeqa
one can obtain a relationship between the viscosity of the medium and the constants\footnote{The sum is now restricted over the two species $a=q\spc (\mbox{quark}), g\spc (\mbox{gluon}$) of the QGP.} $C_q$ and $C_g$,  
\beqa
\frac{\eta}{s}=\left(\frac{s_q}{s}\right) N_q C_q+\left(\frac{s_g}{s}\right) N_g C_g\,.
\eeqa
In the above expression $s=s_q+s_g$ is the total entropy and $N_q=\frac{1350\zeta(5)}{14\pi^2}\approx1.031$ and $N_g=\frac{90\zeta(5)}{\pi^4}\approx0.958$ are two constants that come from the phase space integrals over the quark and gluon distribution functions.
The above expression is simply the statement that $\eta_{tot}=\eta_q+\eta_g$.  In order to solve for $C_q$ we need to estimate the relative viscosity from the quark and gluon phases. Following \cite{Bellac} we take $\eta_q\approx 1.70 N_f\spc \eta_g$.  This factor can be understood from the relative strength of the three cross sections relevant to viscosity at leading log order; $gg\to gg$:$qg\to qg$:$qq\to qq$ = $\frac{9}{2}:2:\frac{8}{9}$.

For $N_f=2$ we find $C_q\approx 1.3\eta/s$.  The larger coefficient for quarks intuitively makes sense.  The quarks must be even further out of equilibrium in order to compensate for the quicker relaxation of the gluons.

\end{document}